\documentclass{ifacconf}

\usepackage{graphicx}      
\usepackage{natbib}        
\usepackage{amssymb}
\usepackage{amsmath}
\usepackage{color,soul}

\newtheorem{Assumption}{Assumption}

\newtheorem{Lemma}{Lemma}
\newtheorem{Proposition}{Proposition}

\newtheorem{Definition}{Definition}

\begin{document}
\begin{frontmatter}

\title{Overcoming Output Constraints in Iterative Learning Control Systems by Reference Adaptation} 

\author[First]{Michael Meindl} 
\author[Second]{Fabio Molinari} 
\author[Second]{J\"org Raisch} 
\author[Second]{Thomas Seel}

\thanks[footnoteinfo]{We acknowledge support of the Deutsche F\"orderungsgemeinschaft (DFG, German Research Foundation) under Germany's Excellence Strategy -- EXC-2002/1 -- Projectnumber 390523135}

\address[First]{Hochschule Karlsruhe -- Technik und Wirtschaft, 
	Germany (e-mail: meindlmichael@web.de).}
\address[Second]{Control Systems Group, Technische Universit\"at Berlin, Germany(e-mail: \{molinari,raisch,seel\}@control.tu-berlin.de)}
\begin{abstract}                
Iterative Learning Control (ILC) schemes can guarantee properties such as asymptotic stability and monotonic error convergence, but do not, in general, ensure adherence to output constraints. The topic of this paper is the design of a reference-adapting ILC (RAILC) scheme, extending an existing ILC system and capable of complying with output constraints. The underlying idea is to scale the reference at every trial by using a conservative estimate of the output's progression. Properties as the monotonic convergence above a threshold and the respect of output constraints are formally proven. Numerical simulations and experimental results reinforce our theoretical results.
\end{abstract}


\end{frontmatter}
\newcommand{\idx}[1]{_{\mathrm{#1}}} 
\newcommand{\mSin}[1]{\text{sin}\left[#1\right]}
\newcommand{\mCos}[1]{\text{cos}\left[#1\right]}
\newcommand{\mTan}[1]{\text{tan}\left[#1\right]}
\newcommand{\mMat}[1]{\mathbf{#1}}
\newcommand{\mVec}[1]{\mathbf{#1}}
\newcommand{\mEquiv}{\hspace{0.5cm}\Longleftrightarrow\hspace{0.5cm}}
\newcommand{\mAbs}[1]{\vert {#1} \vert}
\newcommand{\mTwoNorm}[1]{\left\vert\left\vert {#1} \right\vert\right\vert_{2}}
\newcommand{\mMin}[1]{\text{min}\left[{#1}\right]}
\newcommand{\mMax}[1]{\text{max}\left\{{#1}\right\}}
\newcommand{\mInfNorm}[1]{\left\vert\left\vert {#1} \right\vert\right\vert_\infty}
\newcommand{\mOneNorm}[1]{\left\vert\left\vert {#1} \right\vert\right\vert\idx{1}}
\newcommand{\mVert}{\hspace{1cm}\vert \hspace{0.5cm}}
\newcommand{\mPNorm}[1]{\left\vert\left\vert {#1} \right\vert\right\vert_p}
\newcommand{\mRealNumbers}{\rm I\!R}
\newcommand{\mNaturalNumbers}{\rm I\!N}
\newcommand{\mNaturalNumbersZero}{\rm  I\!N\idx{0}}
\newcommand{\mPosRealNumbers}{\mRealNumbers_{>0}}
\newcommand{\mPosRealNumbersZero}{\mRealNumbers_{\geq 0}}
\newcommand{\mNorm}[1]{\left\vert\left\vert {#1} \right\vert\right\vert}
\newcommand{\mYj}{\mVec{y}_j}
\newcommand{\mP}{\mMat{P}}
\newcommand{\mUj}{\mVec{u}_j}
\newcommand{\mD}{\mVec{d}}
\newcommand{\mYmax}{y\idx{max}}
\newcommand{\mR}{\mVec{r}}
\newcommand{\mL}{\mMat{L}}
\newcommand{\mQ}{\mMat{Q}}
\newcommand{\mUjpone}{\mVec{u}_{j+1}}
\newcommand{\mEj}{\mVec{e}_j}
\newcommand{\mEjpone}{\mVec{e}_{j+1}}
\newcommand{\mRj}{{\mVec{r}}_j}
\newcommand{\mUzero}{\mVec{u}\idx{0}}
\newcommand{\mYzero}{\mVec{y}\idx{0}}
\newcommand{\mYjpone}{\mVec{y}_{j+1}}
\newcommand{\mEinf}{\mVec{e}_\infty}
\newcommand{\mPf}{\mathbf{p}\left(\mUj,\mD\right)}
\newcommand{\mLf}{\mathbf{l}\left(\mUj,\mEj\right)}
\newcommand{\mYinf}{\mVec{y}_\infty}
\newcommand{\mRinf}{\mVec{r}_\infty}
\newcommand{\mUinf}{\mVec{u}_\infty}
\newcommand{\mEmaxEstimate}{\hat{e}_\infty}
\newcommand{\mAj}{a_j}
\newcommand{\mAinf}{a_\infty}
\newcommand{\mAjpone}{a_{j+1}}
\newcommand{\mI}{\mMat{I}}
\newcommand{\mRbound}{\hat{r}}
\newcommand{\mEjTilde}{\mVec{e}_j}
\newcommand{\mForAllJ}{\forall j \in\mNaturalNumbersZero,}
\newcommand{\mYjT}{\mVec{y}_j}
\newcommand{\mYjponeT}{\mVec{y}_{j+1}}
\newcommand{\mUjT}{\mVec{u}_j}
\newcommand{\mUjponeT}{\mVec{u}_{j+1}}
\newcommand{\mMaxSV}{\tilde{\sigma}}
\newcommand{\mEpsi}{\epsilon}
\newcommand{\mGammaTwo}{\gamma\idx{2}}
\newcommand{\mEpsiHat}{\hat{\epsilon}}
\newcommand{\mAjT}{\tilde{a}_j}

\newcommand{\mTheta}{\Theta}
\newcommand{\mThetad}{\dot{\Theta}}
\newcommand{\mS}{s}
\newcommand{\mSd}{\dot{s}}
\newcommand{\mU}{u}
\newcommand{\mZ}{\mVec{z}}
\newcommand{\mZd}{\dot{\mZ}}
\newcommand{\mFi}[1]{f\idx{#1}\left(\mZ\right)}
\newcommand{\mGi}[1]{g\idx{#1}\left(\mZ\right)}
\newcommand{\mV}{\mVec{v}}
\newcommand{\mSVU}{\mVec{u}}
\newcommand{\mSVV}{\mVec{v}}
\newcommand{\mEjTotal}{\mVec{e}_j}
\newcommand{\mEjponeTotal}{\mVec{e}_{j+1}}
\newcommand{\mEinfTotal}{\hat{\mVec{e}}_\infty}
\newcommand{\mXp}{\mVec{x}_{j+1}}
\newcommand{\mX}{\mVec{x}_{j}}
\newcommand{\mXzero}{\mVec{x}\idx{0}}
\newcommand{\mA}{\mMat{A}}
\newcommand{\mAMatj}{\mMat{A}_j}
\newcommand{\mB}{\mMat{B}}
\newcommand{\mb}{\mVec{b}}
\newcommand{\mJT}{j\idx{T}}
\newcommand{\mXUpper}{\tilde{\mVec{x}}_j}
\newcommand{\mEjTTotal}{\hat{\mVec{e}}_{\mJT}}
\newcommand{\mEzeroTotal}{\hat{\mVec{e}}\idx{0}}
\newcommand{\mMaxGamma}{\tilde{\gamma}}
\newcommand{\mRAILCEquations}{(\ref{adapted_update_law})-(\ref{aj_definition})}
\newcommand{\mILCEquations}{(\ref{system_dynamics}-\ref{error_definition})}
\newcommand{\mDEstimate}{\hat{d}}
\newcommand{\mILCMatrix}{\mP\mQ\left(\mI-\mL\mP\right)\mP^{-1}}
\newcommand{\mConstraintFunction}{\mYmax-\mInfNorm{\mRj}}
\newcommand{\mConstraintLimit}{\underline{\varrho}_j}
\newcommand{\mDisturbanceFunction}{\epsilon_j}
\newcommand{\mDisturbanceLimit}{\overline{\epsilon}}
\newcommand{\mAssumptionNumbers}{\ref{Assumption_y_zero}-\ref{Assumption_disturbance_limit}}
\newcommand{\mGammaInf}{\gamma_{\infty}}

\newcommand{\mSISOSystemDefinition}{\mYj=\mP\mUj+\mD}
\newcommand{\mILCUpdateLaw}{\mUjpone=\mQ\left(\mUj+\mL\mEj\right)}
\newcommand{\mEjDefinition}{\mEj=\mR-\mYj}
\newcommand{\mYConstraint}{\mInfNorm{\mYj}\overset{!}{<}\mYmax}

\newcommand{\mADef}{\mA\in\mRealNumbers^{N\times N}}
\newcommand{\mBDef}{\mB\in\mRealNumbers^{N\times N}}
\newcommand{\mVDef}{\mV\in\mRealNumbers^{N\times N}}
\newcommand{\mY}{a}

\newcommand{\mVT}{\mV^T}
\newcommand{\mBT}{\mB^T}
\newcommand{\mAT}{\mA^T}
\newcommand{\mTrans}[1]{{#1}^T}

\section{Introduction}
Iterative Learning Control (ILC) is a control scheme
suitable
for systems operating in a repetitive manner.
ILC tracks a desired reference
and aims at improving its accuracy
from repetition to repetition by exploiting
the information of previous trials,
see \cite{bristow2006survey} for a survey.
ILC can achieve better performance than conventional feedback systems as the latter does not exploit available information from previous trials, see, e.g., \cite{Moore1992Jul}. ILC finds much appeal in fields such as robotics, manufacturing, and biomedical applications 
due to the possibility of repeating trials; such fields 
have greatly benefited from the 
implementation of ILC systems, see, e.g. \cite{Elci2002Aug}, \cite{Seel2011Jan}, \cite{Rogers2016Sep},
and \cite{Pandit1999May}.

\begin{figure}[!ht]
	\label{fig:contraintViolation}
	\begin{centering}
		\includegraphics[width=3.95cm, trim={1.8cm 1.175cm 1cm 0cm}, clip]{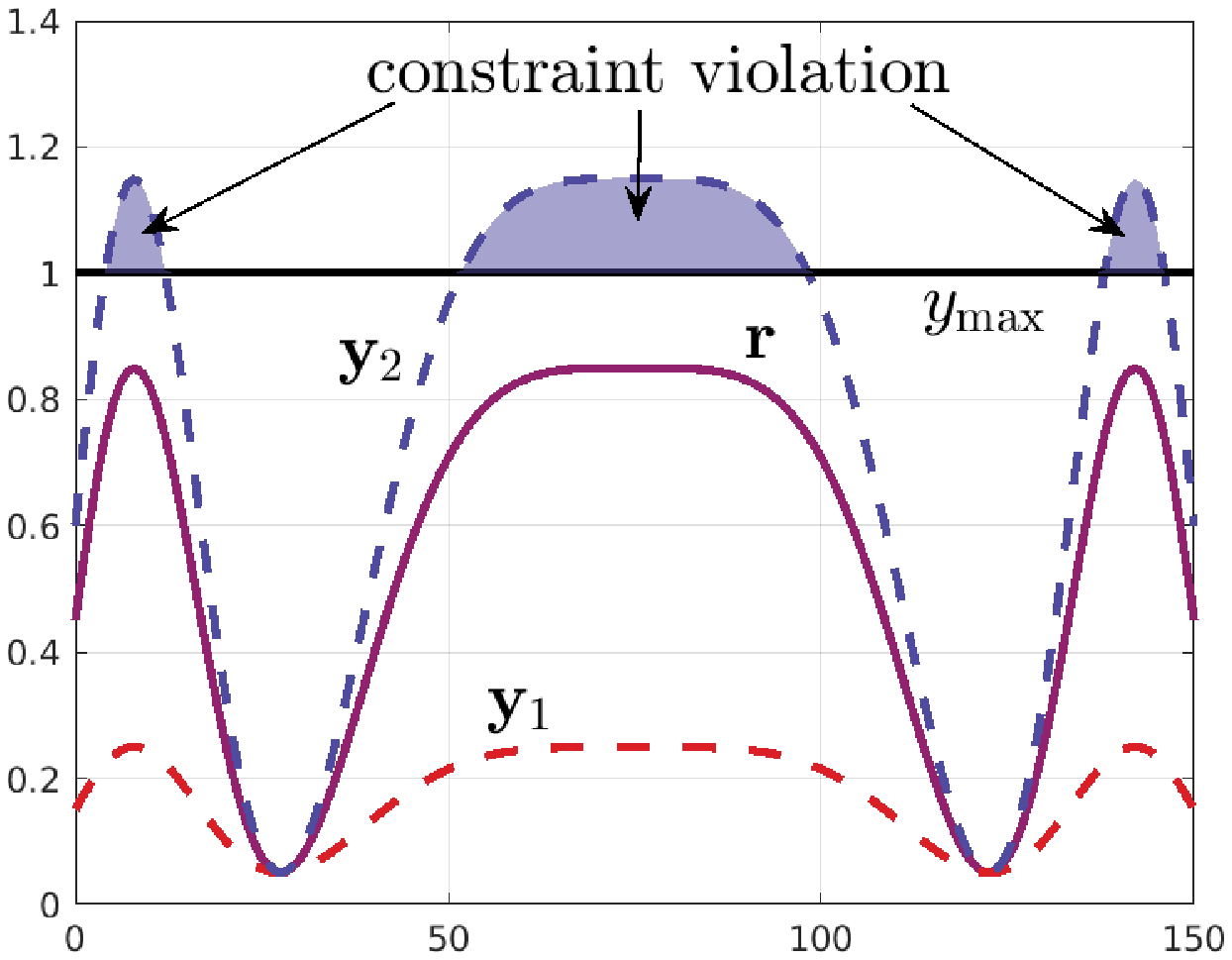}
		\includegraphics[width=4.45cm]{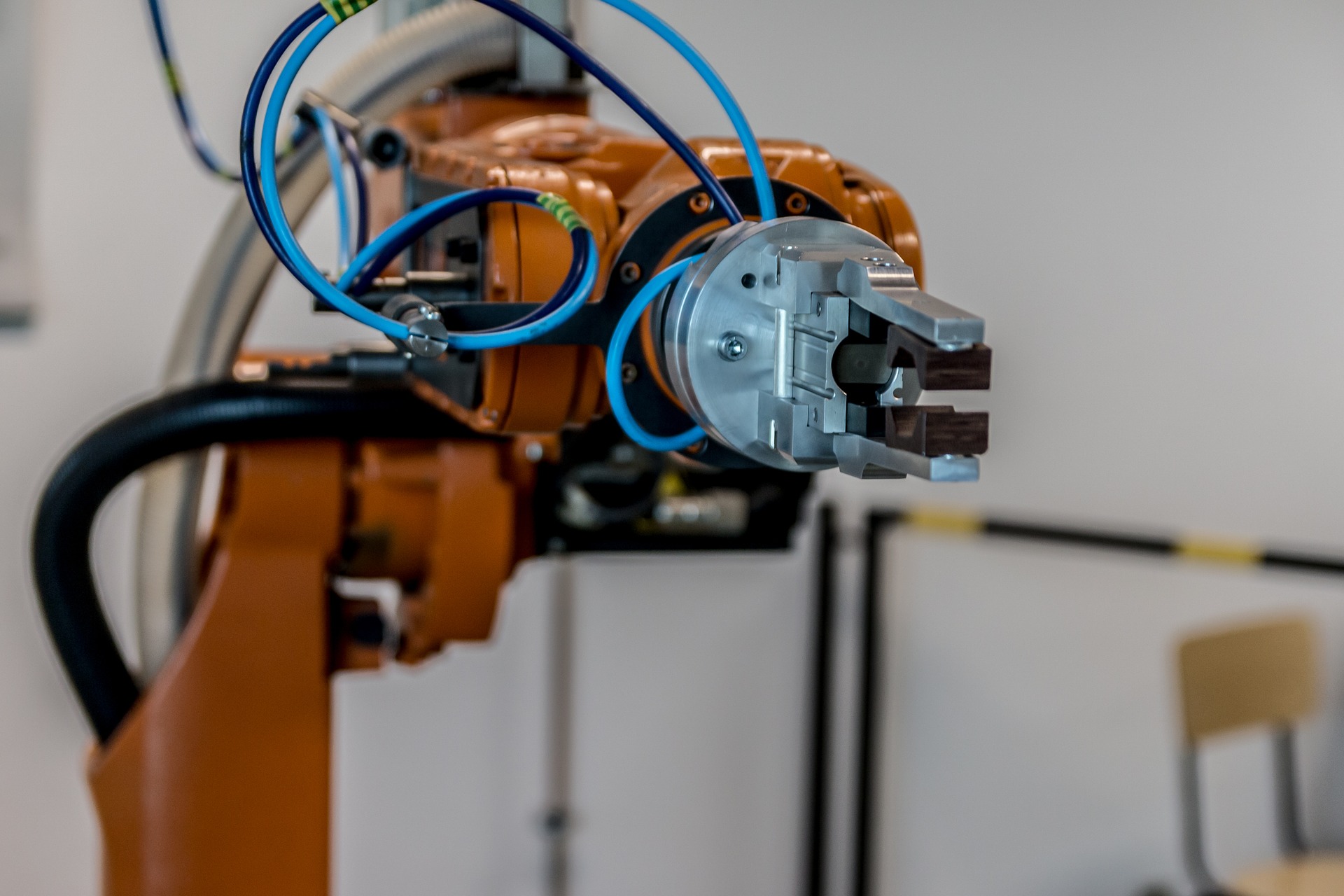}
	\end{centering}
	\caption{Illustration of the output-constrained ILC problem and a robotic manipulator restricted by output constraints}
\end{figure}

A relevant
class of systems 
for ILC applications are those subject to output constraints. 
A robotic manipulator, for example, is commonly restricted when it comes to possible positions and paths; violating these constraints may potentially damage the system. 
In general, ILC systems can guarantee
monotonic error convergence, meaning 
that the difference between
the output and the desired trajectory
decreases in a suitable norm at every trial.
%
However, compliance with 
output constraints is, in general, not guaranteed.
For example, Fig.~\ref{fig:contraintViolation} 
depicts
an ILC system learning the desired trajectory 
$\mR$ while being constrained by the maximum value $\mYmax$. 
Although the system is monotonically convergent, 
the output constraints are violated during the second trial.
This paper addresses this problem
and proposes a modular solution extending 
the ILC system, thus ensuring 
compliance with the output constraints 
while maintaining monotonic error convergence.

Output-constrained ILC systems 
have already been investigated elsewhere
and can be classified in two different categories.
In the first class, 
novel update laws based on constrained optimization have been introduced. For example, \cite{Jin.2014} propose an update-law based on a high-dimensional 
and constrained optimization problem.
This requires accurate knowledge of the plant dynamics
to ensure compliance with the output constraints in time-varying ILC systems. 
In the second class, 
methods like the one proposed by 
\cite{Sebastian.2018} 
combine ILC with additional feedback 
control to handle linear, 
time-varying, multi-input multi-output 
systems with input and output constraints.  \cite{Sebastian.2018Oct} validate this 
approach for a robotic manipulator where an underlying feedback controller 
is implemented to avoid the violation of constraints.

The approach in this paper is to update the reference trajectory for each trial
based on a low-dimensional optimization problem.
It does not require accurate knowledge 
of the plant dynamics
and ensures compliance with the output constraints 
while maintaining the existing ILC system's monotonic 
error convergence. 
In contrast to previous work,
the proposed RAILC scheme is a modular extension
requiring neither high-dimensional optimization 
nor the design of an underlying feedback controller.
The proposed method is validated
using a two-wheeled inverted pendulum robot (TWIPR),
whose task is
to learn a highly accurate motion 
in the presence of output constraints.

The remainder of this paper is structured as follows. Section 2 introduces notation. Section 3 briefly summarizes the structure of conventional ILC system consisting of a linear plant, learning function, and Q-filter. Section 4 formulates the problem. Section 5 introduces a RAILC algorithm and discusses its properties formally. Section 6 presents a TWIPR and validates the RAILC scheme in both simulation and experiment. The simulation results of the RAILC algorithm are also compared to those of the conventional ILC.
Finally, Section 7 gives concluding remarks.

\section{Notation}
Let $\mNaturalNumbersZero$ and $\mNaturalNumbers$ denote the set of nonnegative, respectively positive, integers, $\mRealNumbers$ the set of real numbers and $\mPosRealNumbers$, respectively $\mPosRealNumbersZero$, the set of positive, respectively nonnegative, real numbers. Vectors are in bold type and lower-case letters, e.g., $\mVec{v}$. Matrices are in bold type and upper-case letters, e.g., $\mMat{A}$.
Matrix $\mMat{A}^T$ denotes the transpose of the matrix $\mMat{A}$. Indices of vectors are used to denote the ILC trial, e.g., $\mVec{v}_j$ denotes the vector $\mVec{v}$ on the $j^{\mathrm{th}}$ ILC trial. Let $\mNorm{\mVec{v}}$ denote a norm of the vector $\mVec{v}$, and $\mNorm{\mMat{A}}$ the corresponding, induced matrix norm of the matrix $\mMat{\mMat{A}}$. Particular examples are the Euclidean norm, denoted by $\mTwoNorm{\boldsymbol{\cdot}}$, and the infinite norm, denoted by $\mInfNorm{\boldsymbol{\cdot}}$. The spectral radius of a square matrix $\mMat{A}$ is denoted by $\rho\left(\mMat{A}\right)$. The maximum singular value of matrix $\mMat{A}$ is denoted by $\overline{\sigma}\left(\mMat{A}\right)$.

\section{Lifted-System ILC in a Nutshell}
Consider the discrete-time, linear, time-invariant SISO system
at its $j^{\mathrm{th}}$ trial, $j\in\mNaturalNumbersZero$,
\begin{equation}\label{system_dynamics}
\left\{
\begin{array}{cl}
\mathbf{x}_j(n+1) &= \mMat{A}\mVec{x}_j(n)+\mMat{B}u_j(n)+d(n)\\
y_j(n) &= \mMat{C}\mVec{x}_j(n)
\end{array}\right. \,,
\end{equation}
where $n\in\left\{1,...,N \right\} $ is the sample index 
and $N\in\mNaturalNumbers$ is the numbers of samples in each trial.  Let, $\forall n \in \left\{1,...,N\right\}$, $\mVec{x}(n)\in\mRealNumbers^k$  denote the $k$ dimensional state vector, $u(n)\in\mRealNumbers$ the input variable, $y(n)\in\mRealNumbers$  the output variable and $d(n)\in\mRealNumbers$ an unknown disturbance, which is the same for all trials. Let $\mMat{A}\in\mRealNumbers^{k \times k}$ denote the system matrix, $\mMat{B}\in\mRealNumbers^{k\times 1}$ the input matrix, and $\mMat{C}\in\mRealNumbers^{1\times k}$ the output matrix.

System (\ref{system_dynamics}) can be rewritten in its so-called lifted form, i.e.,
\begin{equation}\label{lifted_system_dynamics}
\mYj = \mMat{P}\mUj + \mD\,,
\end{equation}
where, at each iteration $j\in\mNaturalNumbersZero$,
\begin{subequations}
	\begin{flalign}
	\mUj &= \begin{bmatrix}
	u_j(0) &u_j(1) &\dots &u_j(N-1)
	\end{bmatrix}^T \\
	\mYj &= \begin{bmatrix}
	y_j(m) & y_j(m+1) & \dots& y_j(m+N-1)
	\end{bmatrix}^T \\
	\mD &= \begin{bmatrix}
	d(m) & d(m+1)& \dots& d(m+N-1)
	\end{bmatrix}^T
	\end{flalign}
\end{subequations}
and $m\in\mNaturalNumbersZero$ denotes the system's relative degree, see \cite{bristow2006survey}. For system (\ref{system_dynamics}) the parameters $p_{ij}\in\mRealNumbers$ of the plant matrix $\mP$ are given by
\begin{equation}
\forall i,j\in\mNaturalNumbers\qquad	p_{ij} = \left\{\begin{array}{cl}
\mMat{C}\mMat{A}^{i-j+m-1}\mMat{B} & \forall i\geq j \\
0 & \forall i < j
\end{array}\right.\,.
\end{equation}

The goal of an ILC system is to have output $\mYj$ follow a desired trajectory $\mR \in \mRealNumbers^N$. 
To this end, 
a learning matrix, namely $\mL\in\mRealNumbers^{N\times N}$,
and
a Q-filter, namely $\mQ\in\mRealNumbers^{N\times N}$,
need to be designed (see, e.g. \cite{bristow2006survey}),
thus leading to the update law
\begin{equation}\label{ilc_eq}
\forall j \in \mNaturalNumbers,\qquad	\mUjpone = \mQ\left(\mUj + \mL\mEj\right)\,,
\end{equation}
where $\mEj\in \mRealNumbers^N$ denotes the error trajectory defined as
\begin{equation}\label{error_definition}
\forall j \in \mNaturalNumbers,\qquad
\mEj  := \mR - \mYj\,.
\end{equation}
Note that the matrices $\mP$, $\mL$ and $\mQ$ are regular.
\begin{Definition}\label{Definition_AS}(\textbf{Asymptotic Stability}, 
	see \cite{bristow2006survey}):
	The ILC system with  dynamics (\ref{lifted_system_dynamics}) and update law (\ref{ilc_eq}) is \textit{asymptotically stable} if the 
	limit 
	\begin{align}
	\mEinf :&= 
	\lim_{j\rightarrow \infty} \mEj\nonumber\\
	&= \left[\mMat{I} - \mP\left[\mMat{I}-\mQ\left(\mMat{I}-\mL\mP\right)\right]^{-1}\mQ\mL\right]\left(\mR-\mD\right)\label{e_inf_definition}
	\end{align}
	uniquely exists. In what follows, let
	$\mEinf$ be named \emph{residual error}.
\end{Definition}
Note that,
if $\mQ=\mI$, 
by (\ref{e_inf_definition}),
$\mEinf=0$.
\begin{Proposition}\cite[Theorem 1]{bristow2006survey}s
	The ILC system with dynamics (\ref{lifted_system_dynamics}) and update law (\ref{ilc_eq}) is asymptotically stable if and only if
	\begin{equation}\label{condition_AS}
	\rho\left(\mQ\left(\mMat{I}-\mL\mP\right)\right) < 1\,.
	\end{equation}
	\begin{pf}
		See \cite[Theorem 1]{bristow2006survey}.
	\end{pf}
\end{Proposition}
Despite asymptotic stability,
the error trajectory 
can take arbitrarily large values,
thus generating large learning transients.
The concept of \emph{monotonic convergence}
narrows the set of possible error trajectories.
\begin{Definition}(\textbf{Monotonic Convergence}, see \cite{bristow2006survey}): \label{Definition_MC}
	The system composed of
	plant~(\ref{lifted_system_dynamics}) 
	and
	update law~(\ref{ilc_eq}) 
	is
	\textit{monotonically convergent} 
	under a given norm $\mNorm{\boldsymbol{\cdot}}$ 
	if
	\begin{equation}\label{mc_conventional_ILC}
	\mForAllJ\quad \mNorm{\mEinf-\mEjpone} \leq \gamma\mNorm{\mEinf-\mEj}\,,
	\end{equation}
	where  $\gamma\in[0,1)$ is the convergence rate.
\end{Definition}
\begin{Proposition}
	The system~(\ref{lifted_system_dynamics})-(\ref{ilc_eq})  
	is monotonically convergent under a given norm $\mNorm{\boldsymbol{\cdot}}$
	if
	\begin{equation}\label{condition_MC}
	\gamma := \mNorm{\mP\mQ\left(\mI-\mL\mP\right)\mP^{-1}} < 1\,.
	\end{equation}
	\begin{pf}
		See \cite{bristow2006survey}.
	\end{pf}
\end{Proposition}
Monotonic convergence is traditionally
verified for the spectral and the infinite norms, i.e., 
respectively,
\begin{align}
\mGammaTwo &:= \overline{\sigma}\left(\mP\mQ\left(\mI-\mL\mP\right)\mP^{-1} \right)\\
\intertext{and}
\label{gamma_inf_definition}
\mGammaInf &:= \mInfNorm{\mP\mQ\left(\mI-\mL\mP\right)\mP^{-1}}\,.
\end{align}
\begin{Definition}(\textbf{Monotonic convergence above a}\linebreak \textbf{threshold}, see \cite{Seel2017Mar}): \label{Definition_MC_Threshold}
	System~(\ref{lifted_system_dynamics})-(\ref{ilc_eq})  
	is \textit{monotonically convergent above a threshold} $\kappa\in\mRealNumbers_{\geq 0}$ under a given norm $\mNorm{\boldsymbol{\cdot}}$
	if
	\begin{equation}
	\forall j\in\mNaturalNumbersZero:\mNorm{\mEj}\geq \kappa \quad \implies
	\mNorm{\mEjpone}\leq \mNorm{\mEj}\,.
	\end{equation}
\end{Definition}
\begin{Proposition}\cite[Theorem 1]{Seel2017Mar}
	System~(\ref{lifted_system_dynamics})-(\ref{ilc_eq}) 
	is monotonically convergent above the threshold $\epsilon = \mNorm{\left(\mI-\mP\mQ\mP^{-1}\right)\left(\mR-\mD\right)}$ under a given norm $\mNorm{\boldsymbol{\cdot}}$ if
	\begin{equation}
	\gamma  < 1\,.
	\end{equation}	
	\begin{pf}
		See \cite[Theorem 1]{Seel2017Mar}.
	\end{pf}
\end{Proposition}
By Propositions 1-3, 
to ensure properties as
asymptotic stability and
monotonic convergence,
standard design
methods, see, e.g., \cite{bristow2006survey}, can be employed.
Commonly, the Q-filter is chosen as a 
low-pass filter so that
frequencies above its bandwidth
are cut-off from the learning,
thus increasing robustness.
%
The learning matrix $\mL$ is 
typically designed by either tuning 
the parameters of a PD-function, 
applying $\mathcal{H}_\infty$ optimization, 
or by solving a quadratic optimal problem.

\section{Problem Formulation}
Consider system~(\ref{lifted_system_dynamics})-(\ref{ilc_eq})
with a desired trajectory $\mR$. 
Both the Q-filter and the learning matrix are 
designed such that asymptotic stability 
and monotonic error 
convergence are guaranteed.

Additionally, the output $\mathbf{y}_j$ is constrained by an upper bound $\mYmax\in \mPosRealNumbers$, which must not be violated on any trial, i.e., 
\begin{equation}\label{constraint_eq}
\mForAllJ\qquad \mInfNorm{\mYj} < \mYmax\,.
\end{equation}
Despite monotonic convergence, 
the conventional ILC system does not necessarily enforce
output constraints (see Fig.~1). 
We propose to extend the designed ILC in a modular fashion,
so that output constraints are satisfied
and monotonic error convergence maintained.

To ensure the problem being well-posed, some assumptions are needed.
\begin{Assumption}\label{Assumption_y_zero}
	There is a known initial input trajectory $\mUzero$ leading to an initial output trajectory $\mYzero$ such that
	\begin{equation}
	\mInfNorm{\mYzero}\leq\mYmax\,.
	\end{equation}
\end{Assumption}
\begin{Assumption}\label{Assumption_r}
	The reference trajectory $\mR$ is chosen such that
	\begin{equation}
	\mInfNorm{\mR}\leq \mYmax\,.
	\end{equation}
\end{Assumption}

\section{Reference-Adapting ILC}
The basic idea of this reference-adapting iterative learning control (RAILC) scheme is to adapt the reference trajectory $\mR$ at each ILC trial to ensure that the output trajectory $\mYj$ does not exceed the maximum value $\mYmax$. To this end, let
\begin{equation}
\mForAllJ \qquad \mRj\in\mRealNumbers^N
\end{equation}
denote the adapted reference trajectory at the $j^{\mathrm{th}}$ trial leading to the adapted update law
\begin{equation}\label{adapted_update_law}
\mForAllJ\qquad \mUjponeT = \mQ\left(\mUjT+\mL\left(\mRj-\mYj\right)\right)\,.
\end{equation}
\begin{figure}[!ht]\label{ilc_graphic}
	\label{fig_experimental_results}
	\begin{center}
		\includegraphics[width=8.4cm, trim={0cm 0cm 7cm 8cm}, clip]{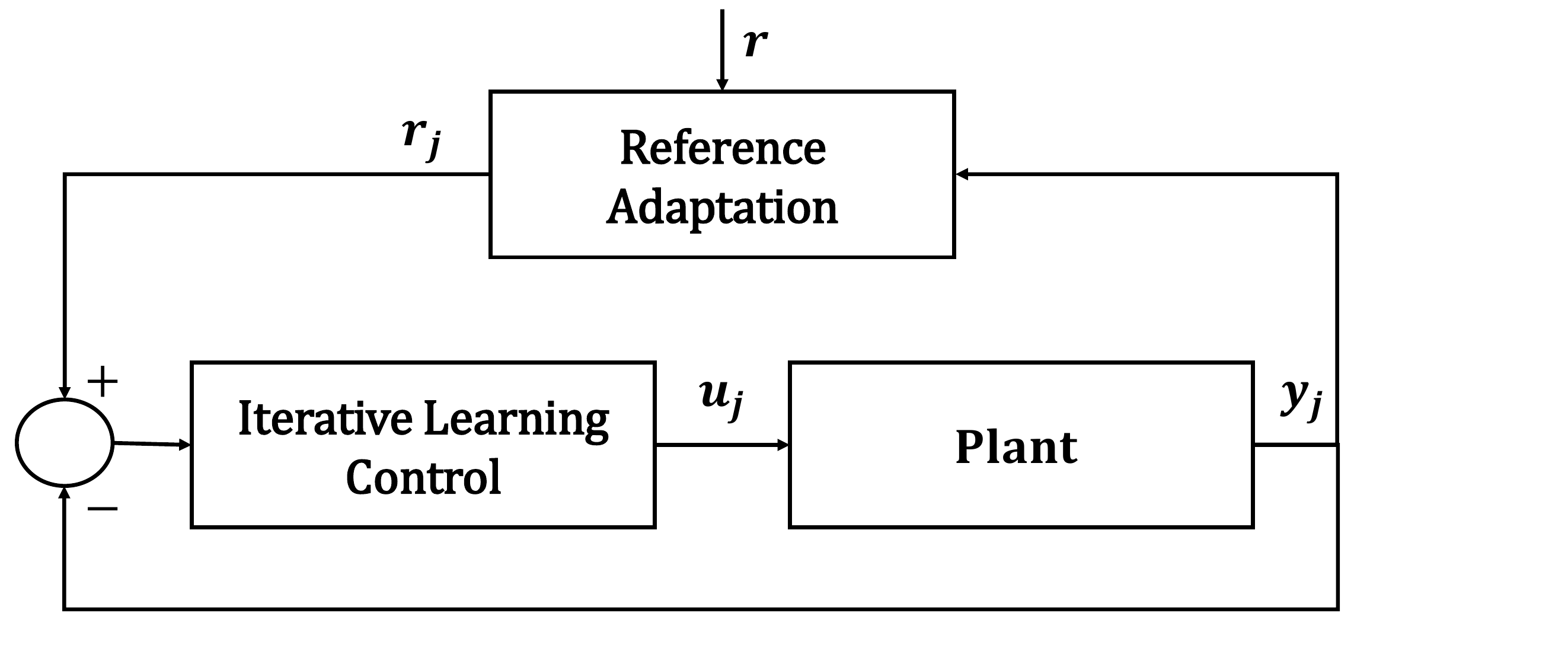}    
		\caption{Block diagram of the RAILC system} 
		\label{fig:bifurcation}
	\end{center}
\end{figure}

Let $\mDisturbanceLimit\in\mRealNumbers$ denote an upper bound such that
\begin{equation}\label{EpsiMax_Definition}
\mDisturbanceLimit \geq \max_{\forall j\in\mNaturalNumbers}{\mInfNorm{\left(\mI-\mP\mQ\mP^{-1}\right)\left(\mRj-\mD\right)}}\,.
\end{equation}

\begin{alg}\label{Algorithm1}
	The adapted reference trajectory $\mRj$ is
	\begin{equation}\label{rj_eq}
	\mForAllJ\qquad	\mRj := \mYjT + \mAj \left(\mR-\mYjT\right)
	\end{equation}
	with
	\begin{subequations}\label{aj_definition}
		\begin{align}
		&\mAj = \max\mAjT \\
		\mathrm{s.t.}\quad &  \mAjT\in[0,1] \\
		&  \mAjT \leq \frac{\mYmax-\mInfNorm{\mYj+\mAjT(\mR-\mYj)}-\mDisturbanceLimit}{\mGammaInf\mInfNorm{\mR-\mYj}}
		\end{align}
	\end{subequations}
\end{alg}

Note that the optimization problem (\ref{aj_definition}) can be efficiently solved using bisection, which makes the algorithm applicable to embedded systems with low computational power.

Consider the upper bound $\mDisturbanceLimit$ required to solve the optimization problem (\ref{aj_definition}). In the trivial case of $\mQ=\mI$, $\mDisturbanceLimit$ can be set to zero. In the case of $\mQ\neq\mI$ an argument similar to the one in \cite{Seel2017Mar} can be applied. In fact,  $\mI-\mP\mQ\mP^{-1}$ commonly takes the form of a high-pass filter. Under the assumption that $\mR$, $\mD$, $\mYzero$ and therefore $\mRj$ 
have spectra well
below the high-pass filter's 
cutoff frequency, the upper bound 
$\mDisturbanceLimit$ can be 
assumed to be a small positive number.

Next, conditions for the existence of a solution to the optimization problem (\ref{aj_definition}) are investigated.
\begin{Proposition}
	For a given $\mDisturbanceLimit$ and $\mYj$, there is a solution to (\ref{aj_definition}) if
	\begin{equation}\label{solution_req}
	\mYmax - \mDisturbanceLimit \geq \mInfNorm{\mYj}\,.
	\end{equation}
\end{Proposition}
\begin{pf}
	$\mAjT=0$ clearly is a feasible solution of (\ref{aj_definition}) if $\mYmax-\mInfNorm{\mYj}-\mDisturbanceLimit \geq 0$; i.e., if (\ref{solution_req}) holds.
\end{pf}
Proposition 4 implies that a solution to the optimization problem (\ref{aj_definition}) is guaranteed to exist if  $$\mYmax-\mInfNorm{\mYj}\geq\mDisturbanceLimit.$$

As previously discussed, $\mDisturbanceLimit$ can be assumed to be a small number. Hence, if $\mYj$ is not ``too close" to $\mYmax$, a solution to (\ref{aj_definition}) exists.

Under this assumption,
we can now focus on the output constraints.

\begin{Proposition}
	If there is a solution to the optimization problem (\ref{aj_definition}), 
	the output constraint~(\ref{constraint_eq}) is guaranteed to hold in the next trial.
\end{Proposition}
\begin{pf}
	If a solution to (\ref{aj_definition}) exists, the following is true
	\begin{equation}\label{a_solution}
	\mAj\mGammaInf\mInfNorm{\mR-\mYj} \leq \mYmax-\mInfNorm{\mRj}-\mDisturbanceLimit\,.
	\end{equation}
	Furthermore, (\ref{rj_eq}) gives
	\begin{equation}
	\mAj\mInfNorm{\mR-\mYj} = \mInfNorm{\mRj-\mYj}\,,
	\end{equation}
	which, incorporated into (\ref{a_solution}), yields
	\begin{equation}\label{P3_eq1}
	\mGammaInf\mInfNorm{\mRj-\mYjT} + \mDisturbanceLimit \leq \mYmax - \mInfNorm{\mRj}\,.
	\end{equation}
	By combining (\ref{adapted_update_law}) 
	and (\ref{lifted_system_dynamics}),
	one obtains
	\begin{equation}
	\mUjpone = \mQ\left(\mI-\mL\mP\right)\mUj + \mQ\mL\left(\mRj-\mD\right)
	\end{equation}
	and
	\begin{multline}
	\mYjpone = \mP\mQ\left(\mI-\mL\mP\right)\mP^{-1}\mYj +\\ \left(\mI-\mP\mQ\mP^{-1}\right)\mD + \mP\mQ\mL\mRj\,,
	\end{multline}
	which leads to
	\begin{multline}\label{eq:1}
	\mRj-\mYjpone = \mP\mQ\left(\mI-\mL\mP\right)\mP^{-1}\left(\mRj-\mYjT\right) +\\ \left(\mI-\mP\mQ\mP^{-1}\right)\left(\mRj-\mD\right)\,.
	\end{multline}
	By applying the inequalities of norms, 
	and by incorporating (\ref{EpsiMax_Definition}) and (\ref{gamma_inf_definition}), one obtains
	\begin{equation}
	\mInfNorm{\mRj-\mYjponeT}\leq \mGammaInf\mInfNorm{\mRj-\mYjT}+\mDisturbanceLimit\,.
	\end{equation}
	Combining this with (\ref{P3_eq1}) yields
	\begin{equation}
	\mInfNorm{\mRj-\mYjponeT}\leq \mYmax - \mInfNorm{\mRj}\,,
	\end{equation}
	which, by adding $\mInfNorm{\mYjpone}$, equals
	\begin{multline}\label{P3_eq3}
	\mInfNorm{\mYjponeT} \leq \mYmax + \mInfNorm{\mYjponeT} - \mInfNorm{\mRj}-\\\mInfNorm{\mRj-\mYjponeT}\,.
	\end{multline}
	Next, consider the norm inequality
	\begin{equation}
	\mInfNorm{\mRj-\mYjponeT} \geq \mInfNorm{\mYjT}-\mInfNorm{\mRj},
	\end{equation}
	equivalently,
	\begin{equation}\label{P3_eq4}
	0 \geq \mInfNorm{\mYjpone}-\mInfNorm{\mRj}-\mInfNorm{\mRj-\mYjponeT}\,.
	\end{equation}
	Combining the latter with (\ref{P3_eq3}) yields
	\begin{equation}
	\mInfNorm{\mYjponeT} \leq \mYmax\,,
	\end{equation}
	which concludes the proof.
\end{pf}

Also, the monotonic convergence properties 
of the RAILC system are investigated.
\begin{Proposition}
	System~(\ref{lifted_system_dynamics})-(\ref{aj_definition}) is monotonically convergent above a threshold $\hat{\kappa}=\frac{\hat{\epsilon}}{1-\hat{\gamma}}$, with
	\begin{equation}\label{eq:epsimaxdef}
	\mEpsiHat := \mNorm{\left(\mI-\mP\mQ\mP^{-1}\right)\left(\mR-\mD\right)}\,,
	\end{equation} 
	under a given norm $\mNorm{\boldsymbol{\cdot}}$
	if
	\begin{equation}\label{MC_eq1}
	\hat{\gamma} := \max_{j\in\mNaturalNumbersZero} \left\{ \mNorm{ \mP\mQ \left( \mI-\mAj\mL\mP \right) \mP^{-1} } \right\} < 1 \,.
	\end{equation}
\end{Proposition}
\begin{pf}
	By combining (\ref{adapted_update_law}) and (\ref{rj_eq}), one obtains
	\begin{equation}
	\mUjpone = \mQ\left(\mI-\mAj\mL\mP\right)\mUj + \mAj\mQ\mL\left(\mR-\mD\right)\,,
	\end{equation}
	which substituted into (\ref{lifted_system_dynamics}) gives
	\begin{multline}
	\mYjpone = \mP\mQ\left(\mI-\mAj\mL\mP\right)\mP^{-1}\mYj +\\ \left(\mI-\mP\mQ\mP^{-1}\right)\mD + \mAj\mP\mQ\mL\mR\,.
	\end{multline}
	Working the latter into  (\ref{error_definition}) leads to
	\begin{multline}\label{eq:explodeEpOne}
	\mEjpone = \mP\mQ\left(\mI-\mAj\mL\mP\right)\mP^{-1}\mEj + \\ \left(\mI-\mP\mQ\mP^{-1}\right)\left(\mR-\mD\right)\,.
	\end{multline}
	Taking the norm, inserting  (\ref{eq:epsimaxdef}) and (\ref{MC_eq1}), and applying the inequalities of norms leads to
	\begin{equation}
	\mNorm{\mEjponeTotal} \leq \hat{\gamma}\mNorm{\mEjTotal} + \mEpsiHat\,,
	\end{equation}
	which, by subtracting $\mNorm{\mEj}$, gives
	\begin{equation}\label{MC_eq4}
	\mNorm{\mEjponeTotal}-\mNorm{\mEjTotal} \leq \left(\hat{\gamma}-1\right)\mNorm{\mEjTotal} +\mEpsiHat\,.
	\end{equation}
	Furthermore, the monotonic convergence of the form
	\begin{equation}
	\mNorm{\mEjponeTotal}\leq \mNorm{\mEjTotal}
	\end{equation}
	equals
	$
	\mNorm{\mEjponeTotal}-\mNorm{\mEjTotal}\leq 0\,,
	$
	which, by (\ref{MC_eq4}), is guaranteed if
	\begin{equation}
	\left(\hat{\gamma}-1\right)\mNorm{\mEjTotal} +\mEpsiHat \leq 0\,.
	\end{equation}
	The latter is equivalent to
	\begin{equation}
	\mNorm{\mEjTotal} \geq \frac{\mEpsiHat}{1-\hat{\gamma}},
	\end{equation}
	thus concluding the proof.
\end{pf}
Recall the argument
that $\mDisturbanceLimit$ can be set to a small number,
if $\mR$, $\mD$, and $\mYzero$ have spectra
below the cutoff frequency of $\mI-\mP\mQ\mP^{-1}$. By (\ref{eq:epsimaxdef}), the same argument applies to $\mEpsiHat$. Regarding (\ref{MC_eq1}), consider the following proposition.
\begin{Proposition}
	System~(\ref{lifted_system_dynamics})-(\ref{aj_definition}) fulfills
	\begin{equation}
	\hat{\gamma}\idx{2} := \max_{j\in\mNaturalNumbers}\left\{\mTwoNorm{\mP\mQ\left(\mI-\mAj\mL\mP\right)\mP^{-1}}\right\} < 1
	\end{equation}
	if
	\begin{equation}
	\mGammaTwo < 1 \quad \textit{and}\quad \mTwoNorm{\mP\mQ\mP^{-1}}\leq 1\,.
	\end{equation}
\end{Proposition}
\begin{pf}
	The proof follows directly from Proposition 8 (see App. B).
\end{pf}
Therefore, the RAILC system is monotonically convergent above a threshold under the Euclidean norm, if the conventional ILC system is monotonically convergent under the Euclidean norm and the mild requirement of $\mTwoNorm{\mP\mQ\mP^{-1}}\leq 1$.

Summarizing the above results, the following statements can be given for the proposed RAILC scheme. 
\begin{itemize}
	\item Under mild assumptions on the spectra of $\mR$, $\mD$ and $\mYzero$, the upper bound $\mDisturbanceLimit$ can be assumed to be a small number.
	\item If $\mYmax - \mDisturbanceLimit \geq\mInfNorm{\mYj}$, a solution to the optimization problem (\ref{aj_definition}) exists.
	\item If a solution to the optimization problem (\ref{aj_definition}) exists, the reference adaption scheme ensures that the output trajectory $\mYjpone$ of the next trial satisfies the output constraints (\ref{constraint_eq}).
	\item If the conventional ILC system is monotonically convergent under the Euclidean norm and $\mTwoNorm{\mP\mQ\mP^{-1}}\leq 1$, the RAILC system is monotonically convergent above a threshold under the Euclidean norm.
	\item The low-dimensional nature of the optimization problem (\ref{aj_definition}) makes the RAILC scheme applicable to systems with low computing power.
	\item The above results were derived under the assumption of $\mP$, $\mL$ and $\mQ$ being regular. Subsequently, the RAILC scheme can also be applied to time-varying, non-causal systems.
\end{itemize}

\section{Experimental Results}
As a demonstrator for the proposed  RAILC, we consider a TWIPR, supposed to perform complex and repeated maneuvers. The robot consists of the pendulum body housing the main electronics, i.e., a microcomputer, inertial measurement units, motors and accumulator. Wheels are mounted onto the motors, which combined with the robot's body create an inverted pendulum. Aiming at keeping the vehicle in an upright position, the system is clearly unstable, thus requiring feedback control.
Two aspects encourage the use of ILC:
the possibility of repeating trials and the lack of precise knowledge regarding the dynamics and the disturbances.

First, let us briefly introduce the TWIPR's dynamics. We consider a TWIPR moving along a straight line. The motor torque is the input variable and is denoted by $u\in\mRealNumbers$. Let $\mTheta\in\mRealNumbers$ denote the pendulum's pitch angle and let $s\in\mRealNumbers$ denote the robot's position (see Fig.~3).

\begin{figure}[!ht]\label{twipr_graphic}
	\label{fig_experimental_results}
	\begin{center}
		\includegraphics[width=5.9cm, trim={0.5cm 2.6cm, 1.7cm, 3cm}, clip]{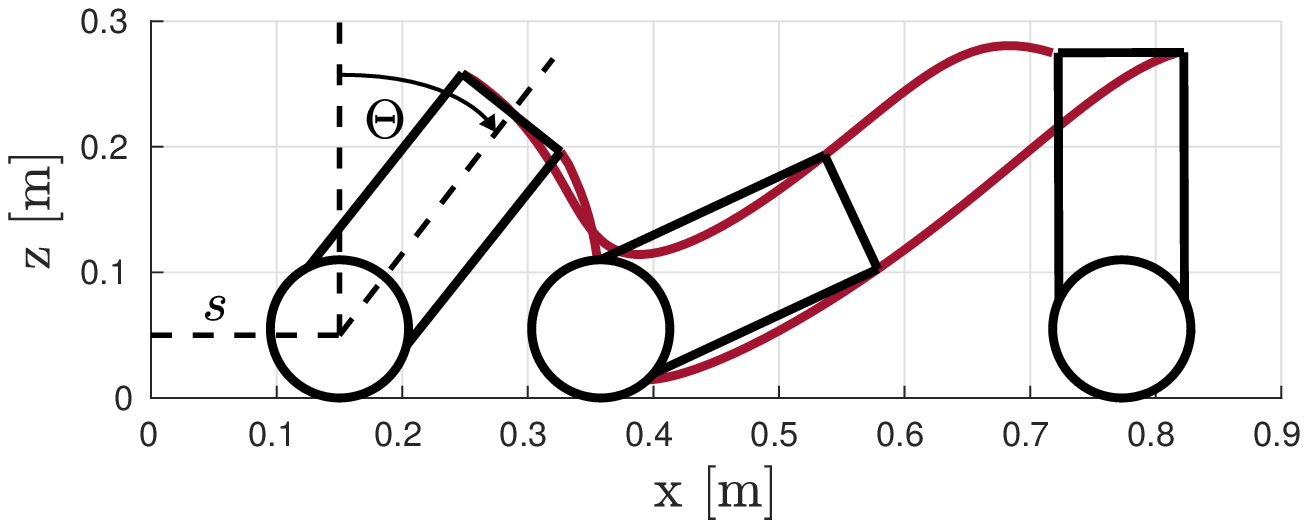}    
		\includegraphics[width=2.5cm, trim={18cm 8.5cm 18cm 8cm}, clip]{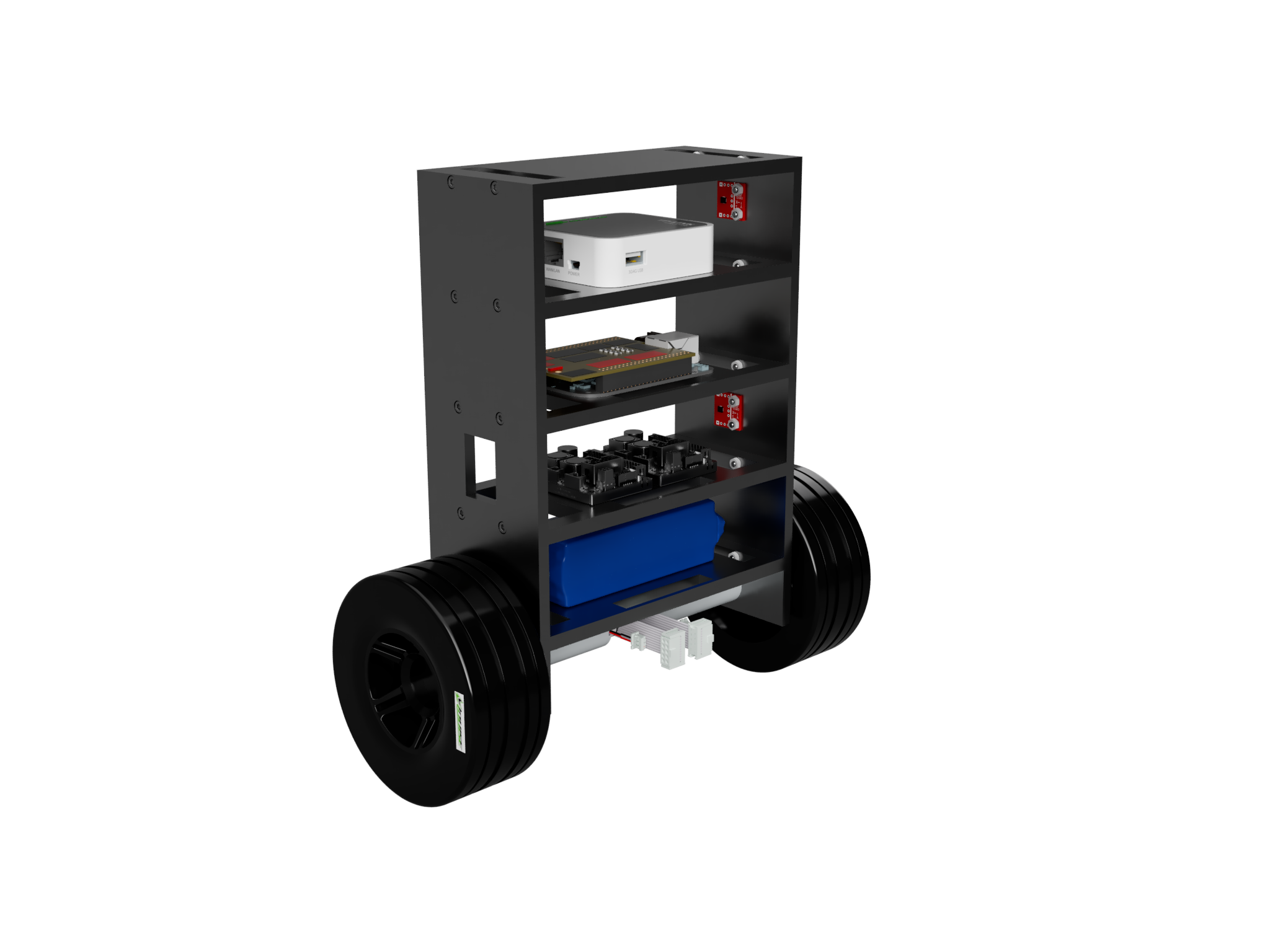}
		\caption{The two-wheeled inverted pendulum robot} 
		\label{fig:bifurcation}
	\end{center}
\end{figure}
The state vector $\mZ\in\mRealNumbers^4$ is defined as
\begin{equation}
\mZ = \left[ \mTheta \hspace{0.2cm} \mThetad \hspace{0.2cm} \mS \hspace{0.2cm} \mSd\right]^T\,.
\end{equation}
A thorough derivation of the TWIPR's dynamics can be found in \cite{kim2015dynamic}. It has been successfully employed elsewhere, see, e.g., \cite{Music2018Nov}, whose notation is also employed here.
By this, let the dynamics be
\begin{equation}\label{TWIPR_dynamics}
\forall t\in\mPosRealNumbersZero, \qquad \mZd(t) = f\left(\mZ(t), u(t)\right)\,.
\end{equation}

To stabilize the inverted pendulum, the controller input $u\idx{C}\in\mRealNumbers$ is calculated by a time-discrete feedback controller of the form
\begin{equation}\label{lin_feedback_eq}
u\idx{c} = -\mMat{K}\mZ\,,
\end{equation}
with sampling time $T=0.02\ \mathrm{sec}$. The feedback matrix $\mMat{K}^{1\times 4}$ is designed using pole-placement and the linearised (at the upright equilibrium), discretised form of the dynamics (\ref{TWIPR_dynamics}). 

For demonstration purposes, the reference trajectory
\begin{equation}
r(n) = 1.22\rm{sin}\left(2/3\pi T n\right)
\end{equation}
with $N=150$ samples is chosen, whose lifted form is denoted by $\mR\in\mRealNumbers^N$. To perform this trajectory an ILC system is implemented, which calculates the input variable $u\idx{ILC}\in\mRealNumbers$, which is an additional motor torque leading to the overall input
\begin{equation}
\forall n=1,\dots,N,\quad
u(n) = u\idx{C}(n) + u\idx{ILC}(n)\,.
\end{equation}
\begin{figure}[!ht]\label{ilc_graphic}
	\label{fig_experimental_results}
	\begin{center}
		\includegraphics[width=8.4cm, trim={0cm 0cm 13.5cm 10.5cm}, clip]{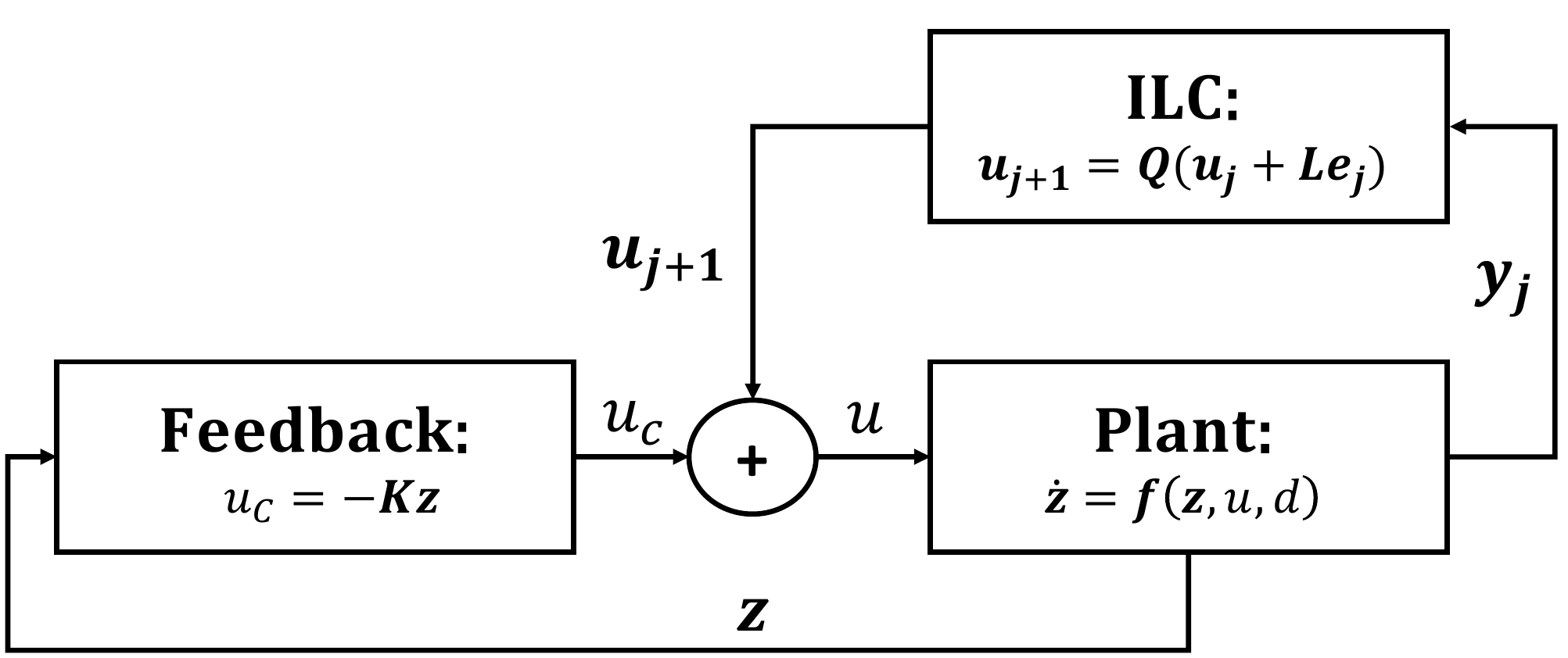}    
		\caption{Block diagram of the ILC system} 
		\label{fig:bifurcation}
	\end{center}
\end{figure}

The lifted form of the ILC's input variable $u\idx{ILC}$ is denoted by $\mUj\in\mRealNumbers^N$, where $j\in\mNaturalNumbers$ denotes the trial index. To update the input trajectory $\mUj$, an ILC law as in (\ref{ilc_eq}),
is applied, where $\mL\in\mRealNumbers^{N\times N}$ denotes the lifted form of the learning function and $\mQ\in\mRealNumbers^{N\times N}$ denotes the lifted form of the Q-filter. To design both these transfer functions, the dynamics of the closed loop plant are linearised leading to the lifted form $\mP\in\mRealNumbers^{N\times N}$ and the linear dynamics (\ref{lifted_system_dynamics}),
where $\mYj\in\mRealNumbers^N$ denotes the pitch trajectory. The learning matrix and Q-filter are calculated using quadratic optimal design, see \cite{bristow2006survey}, yielding an asymptotically stable, monotonically convergent ILC system with the convergence rates
\begin{equation}\label{gamma_two_twipr}
\mGammaTwo = 0.5
\end{equation}
and
\begin{equation}
\mGammaInf = 0.64\,.
\end{equation}

The ILC is output-constrained as, due to the TWIPR's design, the pitch angle is limited by
\begin{equation}
\mForAllJ\quad \mInfNorm{\mYj}\leq\mYmax = 1.31\ [\mathrm{rad}]\,.
\end{equation}
To demonstrate the impact of this restriction, consider the simulation results of applying the conventional ILC system to the TWIPR's non-linear dynamics, which are displayed in Fig.~4. On the first trial, the output trajectory $\mVec{y}\idx{1}$ violates the constraints characterized by $\mYmax$. 

\begin{figure}[!ht]
	\begin{center}
		\includegraphics[width=\linewidth]{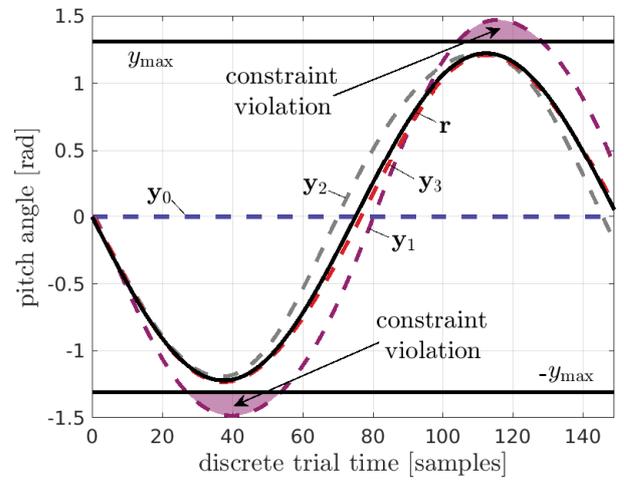}
		\caption{Output Progression of the conventional ILC}
	\end{center}
	\label{simulation_conventional_ILC}
\end{figure}

To ensure that the output constraints are not violated, the RAILC algorithm is applied in simulation first. To meet Assumption \ref{Assumption_y_zero}, the initial input trajectory 
\begin{equation}
\mUzero = \mVec{0}
\end{equation}
is chosen. Assumption \ref{Assumption_r} is also satisfied as the reference trajectory does not exceed the maximum value. Lastly, the upper bounde estimate $\mDisturbanceLimit$ is chosen as
\begin{equation}
\mDisturbanceLimit = 2\mInfNorm{\left(\mI-\mP\mQ\mP^{-1}\right)\mR}\,.
\end{equation}

Results in Fig.~5 show that the RAILC system complies with the output constraints as derived in Proposition 5.
\begin{figure}[!ht]
	\begin{center}
		\includegraphics[width=\linewidth]{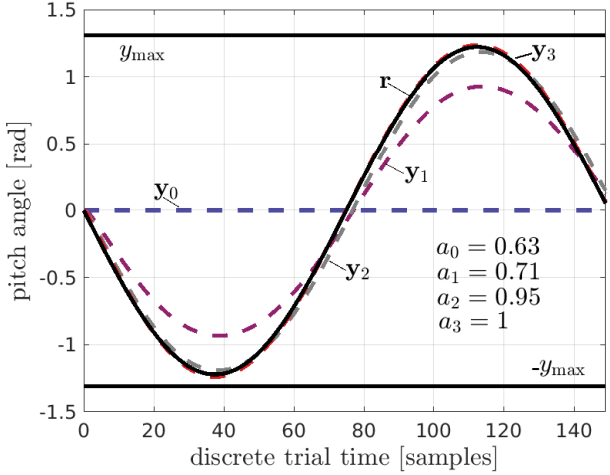}
		\caption{Output Progression of the RAILC in simulation}
	\end{center}
	\label{simulation_RAILC}
\end{figure}

Afterwards, the RAILC algorithm is experimentally validated using the TWIPR. 
Fig.~6 illustrates that the RAILC algorithm also complies with the output constraints when applied experimentally.
\begin{figure}[!ht]
	\begin{center}
		\includegraphics[width=\linewidth]{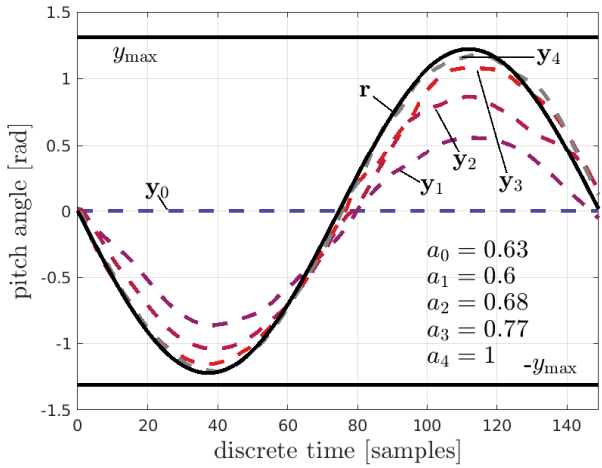}
		\caption{Experimental Output Progression of the RAILC}
	\end{center}
	\label{simulation_RAILC}
\end{figure}

Due to (\ref{gamma_two_twipr}) and (\ref{mc_conventional_ILC}), the conventional ILC system is monotonically convergent. As
\begin{equation}
\mTwoNorm{\mP\mQ\mP^{-1}} = 1
\end{equation}
also fulfills the condition of Proposition 7, the RAILC system is monotonically convergent above a threshold under the Euclidean norm. Fig.~7 shows the progression of the error norms and validates this theoretical finding in both simulation and experiment. Furthermore, in simulation, the RAILC achieves a faster decline of the error norm than the conventional ILC system.

\begin{figure}[!ht]
	\begin{center}
		\includegraphics[width=\linewidth]{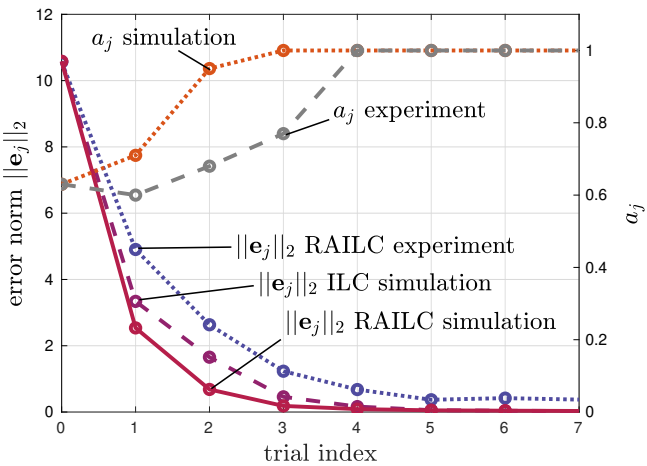}
		\caption{Progression of the error norms $\mNorm{\mEj}$ and $a_j$}
	\end{center}
	\label{simulation_error_progression}
\end{figure}
A solution to the 
reference adaptation problem, 
as discussed in Proposition 1, has been obtained at every trial,
both in simulation and experiment. 
Fig. 7 shows that, in simulation, 
the value of $\mAj$ is monotonically 
increasing and equals one from the third trial onward.
In the experiment, $\mAj$ is not monotonically increasing as there is a  decrease in the first trial. From there onward, $\mAj$ is monotonically increasing and reaches one on the fourth trial.

\section{Conclusion}
We have extended standard
ILC schemes in a modular fashion to cope with output-constrained systems.
By adapting the reference trajectory 
based on a conservative estimate of the 
output progression, the violation of output constraints is avoided. 
Under mild assumptions on disturbances, reference trajectory 
and the output trajectory, the existence of a solution 
is guaranteed.
A condition 
for the RAILC error to monotonically converge
above a threshold was given.
It has been shown that this threshold can be expected to be close to zero. 
The RAILC approach has been applied to a TWIPR, and both simulation and experimental results have been shown.
While the conventional ILC system violates output constraints, 
the RAILC system complies with them.
Furthermore, the RAILC system exhibits monotonic convergence above a small threshold
in both simulation and experiment. 
Finally, simulation results show that applying RAILC does not slow down the decline of the error norm when compared to the conventional ILC.

Ongoing work investigates conditions for the existence of a solution to the optimization problem (\ref{aj_definition}). Furthermore, we will
study the possibility of using RAILC to increase performance.

\bibliography{Overcoming_Output_Constraints_in_Iterative_Learning_Control_Systems_by_Reference_Adaptation}             
                                              







\appendix
\section{Proposition 8}
\begin{Lemma}
	Given a matrix $\mADef$, a matrix $\mBDef$, and a vector $\mVDef$ with
	\begin{equation}\label{L2:eq1}
	\mTwoNorm{\mV} = 1\,,
	\end{equation}
	if (\textit{sufficient condition})
	\begin{equation}\label{L2:eq3}
	\mTwoNorm{\mB-\mA} < 1\,,
	\end{equation}
	then 
	\begin{equation}\label{L2:eq2}
	\mVT\left(\mBT\mA+\mAT\mB\right)\mV > \mVT\mAT\mA\mV + \mVT\mBT\mB\mV - 1\,.
	\end{equation}
\end{Lemma}
\begin{pf}
	Combining (\ref{L2:eq3}), (\ref{L2:eq1}), and the submultiplicativity of norms gives
	\begin{multline}
	1 > \mTwoNorm{\mB-\mA} = \mTwoNorm{\mB-\mA}\mTwoNorm{\mV} \\\geq \mTwoNorm{\mB\mV-\mA\mV}
	= \sqrt{ \mTrans{\left(\mB\mV-\mA\mV\right)}\left(\mB\mV-\mA\mV\right) }\,,
	\end{multline}
	which, by taking the squares and expanding, leads to (\ref{L2:eq2}), thus, concluding the proof.
\end{pf}
\renewcommand{\mV}{\mVec{w}}

\begin{Proposition}
	Given a matrix $\mADef$, a matrix $\mBDef$,
	and a scalar $\mY\in(0,1]$, if (\textit{sufficient condition})
	\begin{equation}\label{T1:eq2}
	\mTwoNorm{\mB-\mA} < 1\quad \mathit{and} \quad \mTwoNorm{\mB} \leq 1\,,
	\end{equation}	
	then
	\begin{equation}\label{T1:eq1}
	\mTwoNorm{\mB-\mY\mA}<1\,.
	\end{equation}
\end{Proposition}
\begin{pf}
	First, let $\mV\in\mRealNumbers^{N}$ denote a vector such that 
	\begin{equation}\label{T1:eq3}
	\mV = \underset{\mTwoNorm{\mVec{x}}=1}{\mathrm{argmax}}\left\{\mTwoNorm{\left(\mB-\mY\mA\right)\mVec{x}}\right\}\,,
	\end{equation}
	which, according to the definition of induced matrix norms, implies
	\begin{multline}\label{T1:eq5}
	\mTwoNorm{\mB-\mY\mA} = \mTwoNorm{\left(\mB-\mY\mA\right)\mV} 
	= \\
	\sqrt{ \mVT\mBT\mB\mV - \mY\mVT\left(\mBT\mA + \mAT\mB\right)\mV + \mY^2\mVT\mAT\mA\mV}\,.
	\end{multline}
	Lemma 1 further gives
	\begin{equation}
	\mVT\left(\mBT\mA+\mAT\mB\right)\mV > \mVT\mAT\mA\mV + \mVT\mBT\mB\mV - 1\,,
	\end{equation}
	which combined with (\ref{T1:eq5}) leads to
	\begin{multline}\label{T1:eq6}
	\mTwoNorm{\mB-\mY\mA} < \\ \sqrt{(1-\mY)\mVT\mBT\mB\mV + \mY + (\mY^2-\mY)\mVT\mAT\mA\mV}\,. 
	\end{multline}
	Submultiplicativity of norms, (\ref{T1:eq2}), and (\ref{T1:eq3}) give
	\begin{equation}
	\mVT\mBT\mB\mV \leq 1\,,
	\end{equation}
	which combined with
	\begin{equation}
	\mY\in(0,1]~ \implies ~\mY^{2} \leq \mY ~ \implies ~ \mY^{2}-\mY \leq 0
	\end{equation}
	and (\ref{T1:eq6}) leads to
	\begin{equation}
	\mTwoNorm{\mB-\mY\mA} < \sqrt{1-\mY+\mY} = 1\,.
	\end{equation}
	This concludes the proof.
\end{pf}
\end{document}